# Chaotic Sequences for Secure CDMA

Madhekar Suneel

***Abstract:*** **One family each of repetitive and non-repetitive spreading sequences for secure Direct Sequence Spread Spectrum communication are obtained from the logistic map. The repetitive sequences are perfectly balanced and are found to have sufficient linear complexity. Applicability of both the families to CDMA communication is studied by examining their cross-correlation performance. The cross-correlation is found to be comparable to and only rarely worse than that of Gold sequences.**

## I. INTRODUCTION

SEQUENCES derived from chaotic phenomena are actively being considered for spread-spectrum communications [1]-[3]. This paper is concerned with Direct Sequence Spread Spectrum (DSSS) wherein the spreading sequence is directly multiplied with the data sequence to achieve bandwidth spreading [4], [5]. To reconstruct a sequence with linear complexity (LC) of $L$, from an intercepted block of chips, an adversary requires $2L$ contiguous chips of the sequence [6]. Therefore, if the linear complexity of a sequence is larger, the adversary is required to possess larger blocks of the spreading sequence. Further, if the linear complexity is exactly equal to half the length of the sequence, the adversary is required to possess the entire sequence for reconstruction. However, if the adversary is in possession of the entire sequence, there is no need for reconstruction. A larger value of linear complexity does not give any better advantage. Therefore, if a sequence has a linear complexity of half its length, it can be considered secure. Here, one family of chaotic sequences with linear complexity close to half of sequence-length is proposed.

If the sequence does not repeat for consecutive data bits, then reconstruction of a portion of the sequence does not help the adversary in any way. However, synchronization of the receiver to the transmitter's spreading sequence is required. In this paper, one family of such non-repetitive chaotic sequences is also proposed.

The suitability of repetitive and non-repetitive sequences for Code Division Multiple Access (CDMA) is considered by studying their correlation properties.

---

M. Suneel is with PGAD (Defence Research and Development Organization, Ministry of Defence, Government of India), DRDL Complex, Kanchanbagh, Hyderabad-500 058. (email: suneel@ieee.org).

Let $C = \{c_i\}_{i=1}^{N}$ be an $N$-chip sequence and $U = \{u_i\}_{i=1}^{\infty}$ a sequence of indefinite length. Let $U_k^N$ denote the $N$-chip subsequence of $U$ starting with the $k^{\text{th}}$ chip. Let $A_k(C,U)$ and $D_k(C,U)$ respectively denote the number of chip-by-chip agreements and disagreements between $C$ and $U_k^N$. Then for all positive integral values of $k$, the (normalized) acyclic cross-correlation (function) of $C$ and $U$ is defined as

$$\theta(C,U,k) = (A_k - D_k)/N . \qquad (1)$$

Let $B = \{b_i\}_{i=1}^{N}$ be another $N$-chip sequence and $T^k(B)$ denote the sequence obtained by cyclically right-shifting $B$ by $k$ chips. For negative $k$, left-shifting may be assumed. Let $A_k(C,B)$ and $D_k(C,B)$ respectively denote the number of agreements and disagreements between $C$ and $T^k(B)$. Then we define the *(normalized) cyclic cross-correlation (function)* of $C$ and $B$ as

$$\theta(C,B,k) = (A_k - D_k)/N . \qquad (2)$$

$k$ is called the relative phase and $\theta(C,C)$ is called the *auto-correlation* of $C$.

A sequence of even length is *balanced* if the number of ones and zeros is equal. A sequence of odd length is *balanced* if the number of ones is one more than the zeros.

## II. BINARY SEQUENCES FROM THE LOGISTIC MAP

Consider the logistic map [7], [8]

$$x_{n+1} = \lambda x_n (1 - x_n) \qquad (3)$$

where $x \in (0,1)$, $\lambda \in (0,4)$ and $n = 0,1,2,\dots$ It has regions of chaos in the interval $\lambda \in (3.58,4)$ and values close to 4.0 were used for the work presented in this paper. For generating non-repetitive binary sequences from (3), a transformation $\Omega_1 : (0,1) \rightarrow \{0,1\}$ is proposed as follows:

$$\Omega_1(x) = \begin{cases} 1, \text{if } x \geq \tau; \\ 0, \text{if } x < \tau. \end{cases} \qquad (4)$$

where $\tau$ is a threshold value. For balanced nature, in this paper, $\tau$ is chosen as the arithmetic mean of a large number of consecutive values of $x$. Sequences generated by $\Omega_1$ are called $\Omega_1$-sequences in this paper. Multiple-access codes can be generated by using different $\lambda$ and/or $x_0$ for different users.

Let $R = \{r_i\}_{i=1}^{N}$ be a real-valued sequence from a chaotic orbit. Let $R' = \{r_i'\}_{i=1}^{N}$ be the sequence obtained by arranging $R$ in ascending order, i.e. $r_1'$ is the smallest and $r_N'$ is the largest element of $R$. Let $I = \{\iota_i\}_{i=1}^{N}$ be a perfectly balanced binary *initial sequence*. For simplicity sake, in this paper, we assume $I$ to consist of alternate ones and zeros. Let the $j^{\text{th}}$ element of $R$ be the $h^{\text{th}}$ element of $R'$, i.e. $r_h' = r_j$. Then an $N$-chip sequence can be obtained by the transformation

$$\Omega_2(r_j) = \iota_h. \tag{5}$$

In effect, the sequence is obtained by re-arranging $I$ as per the indices of elements of $R$ in $R'$. Since $I$ is chosen as balanced, the resulting sequence is also balanced.

### III. LINEAR COMPLEXITY PERFORMANCE

Linear complexities of a large number of $\Omega_2$-sequences were determined using the Berlekamp-Massey algorithm [9]. Based on the observations, an empirical formula was found for the probability of the linear complexity $L$ of an $N$-chip $\Omega_2$-sequence ($N$ odd) being equal to $c$. It is

$$P(L=c) = \begin{cases} (0.5)^{N-2c+1} & \text{if } c < (N+1)/2; \\ (0.5)^{2c-N} & \text{if } c \ge (N+1)/2. \end{cases} \tag{6}$$

The experimental distribution for $N = 127$ is in Fig. 1.

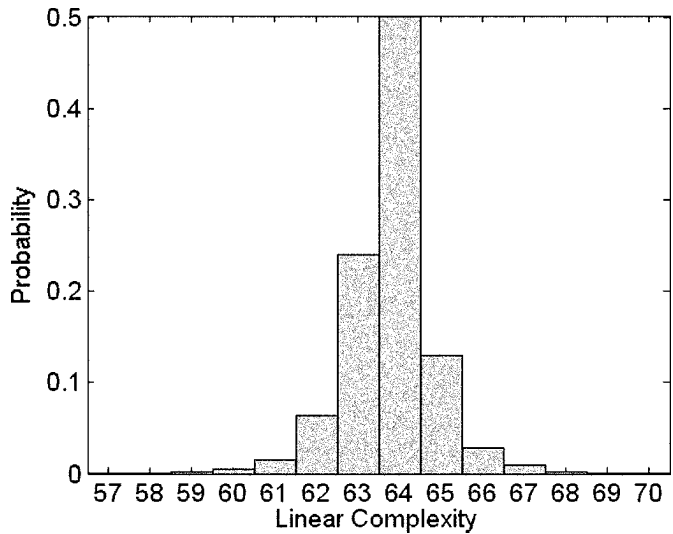

Fig. 1. Experimental LC distribution: 127-chip $\Omega_2$-sequences.

The distribution determined from (6) is shown in Fig. 2.

The equivalence of the experimental and the conjectured distributions is evident.

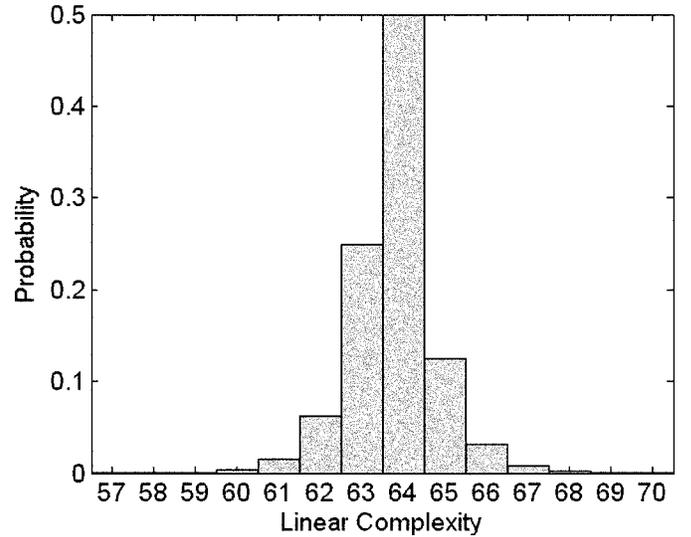

Fig. 2. Conjectured LC distribution: 127-chip $\Omega_2$-sequences.

The empirical formula that was found based on a similar study for even $N$ is

$$P(L=c) = \begin{cases} (0.5)^{N-2c+1} & \text{if } c \le N/2; \\ (0.5)^{2c-N} & \text{if } c > N/2. \end{cases} \tag{7}$$

The experimental distribution for $N = 64$ and that determined by (7) for the same $N$ are shown in Fig. 3 and Fig. 4 respectively. Again, the agreement of conjecture and experiment is evident.

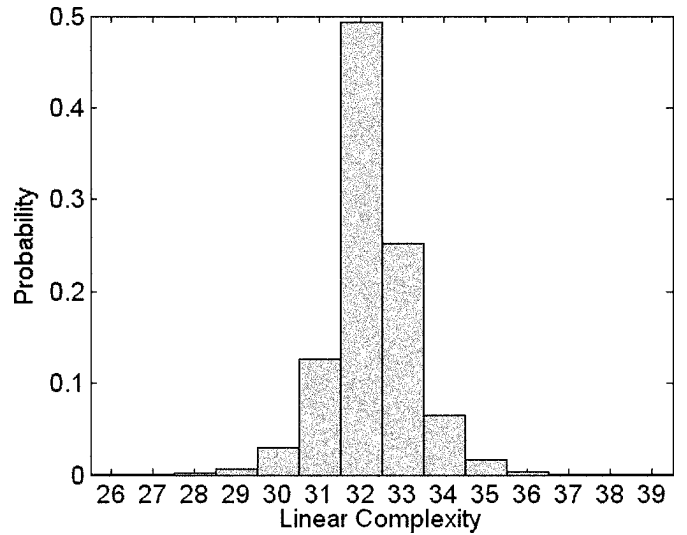

Fig. 3. Experimental LC distribution: 64-chip $\Omega_2$-sequences.

As the linear complexity is very close to half the sequence length, $\Omega_2$-sequences can be considered secure. It is interesting to note that the linear complexity is better than some nonlinear feedforward-logic type shift-register-based sequence generators [10].

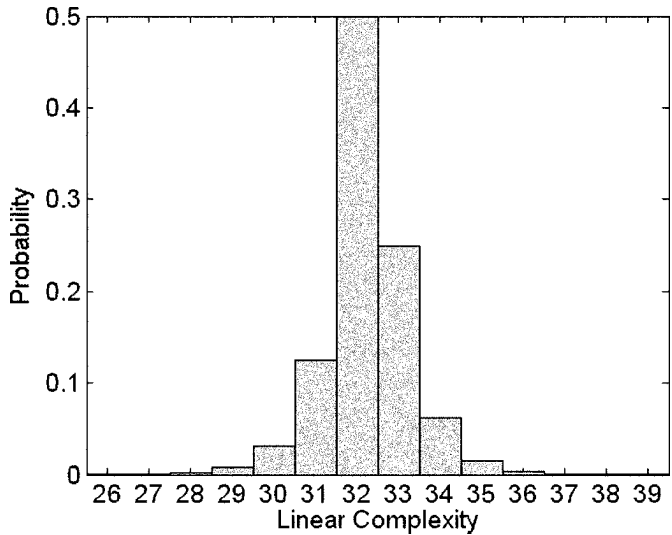

Fig. 4. Conjectured LC distribution: 64-chip $\Omega_2$-sequences.

## IV. CORRELATION PERFORMANCE

### A. $\Omega_1$-Sequences

It can be mathematically proven [11] that the probability of the acyclic cross-correlation of two random 'toss-of-a-coin' sequences being equal to $\Theta$, where

$$\Theta = \begin{cases} 0, \pm \frac{2}{N}, \pm \frac{4}{N}, \ldots, \pm 1 & \text{for even } N, \\ \pm \frac{1}{N}, \pm \frac{3}{N}, \pm \frac{5}{N}, \ldots, \pm 1 & \text{for odd } N. \end{cases} \quad (8)$$

is given by

$$P(\theta = \Theta) = \sqrt{\frac{2}{N\pi}} \, e^{-N\Theta^2/2}. \quad (9)$$

The author has experimentally found the probability distribution of acyclic cross-correlation of $\Omega_1$-sequences to be very close to that given by (9). To demonstrate this point, the probability distribution of acyclic cross-correlation for 1023 chips/bit $\Omega_1$-sequences and that given by (9) for $N = 1023$ are plotted on a single graph in Fig. 5. The equivalence of the probability distributions can readily be seen.

The three-valued cross-correlation of 1023-chip Gold sequences [12], [13] is also shown as dotted vertical lines in Fig. 5 for comparison sake. It can be seen that values of cross-correlation worse than that of Gold sequences occur rarely.

The auto-correlation performance of $\Omega_1$-sequences is not considered here since it is assumed that DSSS communication with non-repetitive sequences is carried out with some external mechanism to synchronize the sequence-generator of the receiver with that of the transmitter.

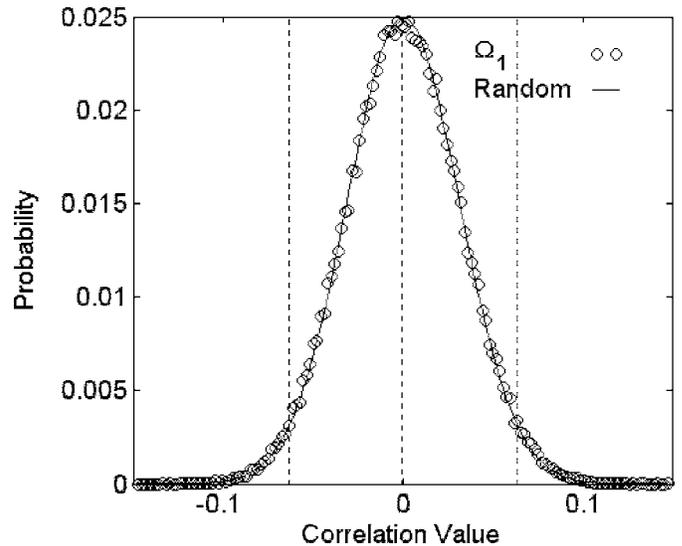

Fig. 5. Probability distribution of acyclic cross-correlation: 1023 chips-per-bit $\Omega_1$-sequences and random sequences. The three valued cross-correlation of 1023-chip Gold sequences is indicated by vertical dotted lines.

### B. $\Omega_2$-Sequences

Let $G$ be any $N$-chip perfectly balanced binary sequence of long length. Let $[H_i]_{i=1}^N$ be a collection of $N$ sequences randomly chosen from the set of all balanced sequences of length $N$. Let $A_k$ and $D_k$ respectively denote the number of agreements and disagreements between $G$ and $H_k$. Then the probability distribution of the quantity $\theta' = (A_k - D_k)/N$ assuming the value $\Theta$ can be shown to be $(k = 1, 2, \ldots, N)$ [11]

$$P(\theta' = \Theta) = 2\sqrt{\frac{2}{N\pi}} \, e^{-N\Theta^2/2}, \quad (10)$$

where

$$\Theta = \begin{cases} 0, \pm \frac{4}{N}, \pm \frac{8}{N}, \ldots, \pm 1 & \text{for even } N, \\ -\left(1 - \frac{2}{N}\right), -\left(1 - \frac{6}{N}\right), \ldots, 1 & \text{for odd } N. \end{cases} \quad (11)$$

The author has experimentally found the probability distribution of cyclic cross-correlation values for a pair of sufficiently long $\Omega_2$-sequences to be very close to that given by (10). As an illustration, the experimental distribution for a pair of 1023-chip $\Omega_2$-sequences is plotted along with the plot of (10) with $N = 1023$ in Fig. 6. The three-valued cross-correlation of 1023-chip Gold sequences [12], [13] is also shown as vertical dotted lines for comparison sake. Again, it can be seen that values worse than the cross-correlation of Gold sequences are rare.

For repetitive sequences, an auto-correlation function closely resembling the $\delta$-function is essential for conventional tracking-loop-based receivers to function. The auto-correlation of several $\Omega_2$-sequences was evaluated and this indeed was the case. As an example, the auto-correlation plot for a 1023-chip $\Omega_2$-sequence is shown in Fig. 7.

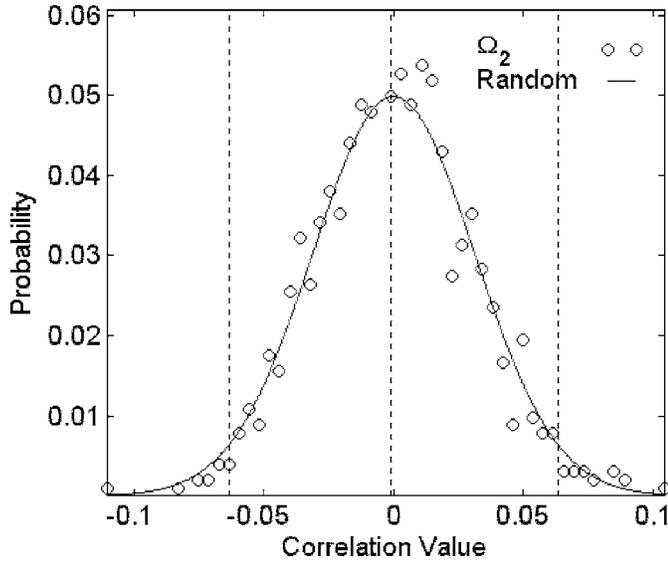

Fig. 6. Probability Distribution of cyclic cross-correlation of 1023-chip $\Omega_2$-sequences and the probability distribution of $\theta'$ (shown as *random* in the legend). The three values of cross-correlation of 1023-chip Gold sequences are indicated as vertical dotted lines.

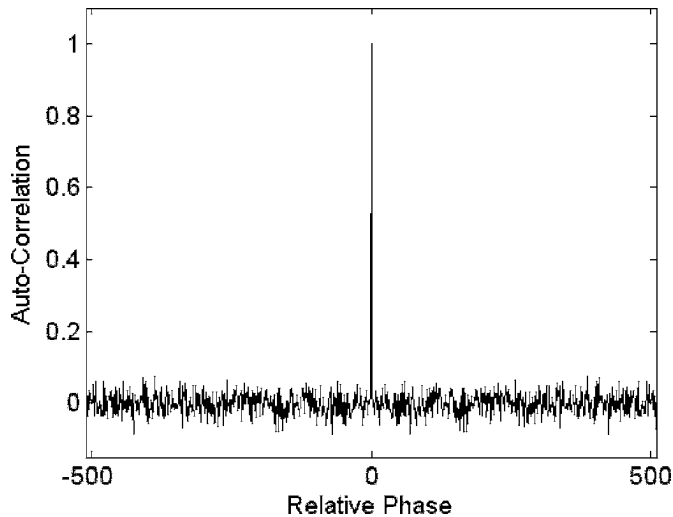

Fig. 7. Auto-correlation: 1023-chip $\Omega_2$-sequence.

## V. CONCLUDING REMARKS

Considering the cross-correlation properties of both $\Omega_1$ and $\Omega_2$-sequences, CDMA communication seems feasible. Since the auto-correlation of $\Omega_2$-sequences is found to have a single prominent peak at zero-phase, asynchronous CDMA communication using these sequences also seems feasible.

Large linear complexity of $\Omega_2$-sequences makes them an attractive candidate for secure asynchronous DSSS/CDMA communications. However, it should be borne in mind that regardless of the linear complexity, if the adversary intercepts one complete period of the repetitive sequence, he can still jam the receiver or gain access to the information content. Therefore, non-repetitive sequences are always more secure than repetitive sequences. On the other hand, asynchronous communication with repetitive sequences is much simpler to implement. Depending on the application, a repetitive-sequence-based system in which the sequence is systematically changed at both the transmitter and the receiver may be considered adequate.

## VI. ACKNOWLEDGMENT

The author would like to thank the Director, Research Centre Imarat for permission to publish. Suggestions of Prof. Soumitro Banerjee (Department of Electrical Engineering, Indian Institute of Technology, Kharagpur) and Soumyajit Mandal (Department of Electrical Engineering and Computer Science, Massachusetts Institute of Technology, Cambridge) have greatly helped in improving this paper. Discussions with R.S. Chandrasekhar (Research Centre Imarat, Hyderabad) have helped with proofs of (9) and (10).